\documentclass[11pt]{article}
\date{}
\usepackage{amsmath,amsfonts,graphicx}

\newtheorem{prop}{Proposition}

\newtheorem{defn}{Definition}

\newtheorem{thm}{Theorem}

\begin{document}

\newcommand{\thmref}[1]{Theorem~\ref{#1}}
\newcommand{\secref}[1]{Sect.~\ref{#1}}
\newcommand{\lemref}[1]{Lemma~\ref{#1}}
\newcommand{\propref}[1]{Proposition~\ref{#1}}
\newcommand{\corref}[1]{Corollary~\ref{#1}}
\newcommand{\remref}[1]{Remark~\ref{#1}}

\newcommand {\beq}{\begin{equation}}
\newcommand {\eeq}{\end{equation}}
\newcommand {\beqa}{\begin{eqnarray}}
\newcommand {\eeqa}{\end{eqnarray}}         
\newcommand {\beqas}{\begin{eqnarray*}}
\newcommand {\eeqas}{\end{eqnarray*}}
\newcommand {\bea}{\begin{array}}
\newcommand {\eea}{\end{array}}
\newcommand {\bds}{\begin{displaymath}}
\newcommand {\eds}{\end{displaymath}}
\newcommand {\nn}{\nonumber}
\newcommand{\no}{\noindent}
\newcommand{\bi}{\bibitem}

\begin{flushright}
\end{flushright}

\baselineskip =18pt

\vskip 1cm

\begin{center}
{\Large\bf The central elements of $E_{q,p}(\widehat{sl_2})_k$ with
the critical level}

\vspace{1cm}

{\normalsize\bf Wenjing Chang$^{a}$\footnote{Email:
wjchang@amss.ac.cn}, Xiang-mao Ding$^{b}$\footnote{Email:
xmding@amss.ac.cn}, Ke Wu$^{a}$\footnote{Email:
wuke@mail.cnu.edu.cn}}

\vskip1cm {\em$^a$ Capital Normal University, Beijing 100048,
People's Republic of China}
\\
{\em$^b$ Institute of Applied Mathematics, Academy of Mathematics
and Systems Science; \\Chinese Academy of Sciences, P.O.Box 2734,
Beijing 100190, People's Republic of China}
\date{}
\end{center}

\begin{abstract}
In this paper we generalize certain results concerning quantum affine algebra $U_{q}(\widehat{sl_{2}})$ at the critical level to the corresponding elliptic case $E_{q,p}(\widehat{sl_2})$. Using the
Wakimoto realization of the algebra $E_{q,p}(\widehat{sl_2})$, we construct the central elements of it at the critical level. It turns out that the so called Drinfeld conjecture originally proposed for Kac-Moody algebras also holds for the elliptic quantum algebras.
\end{abstract}

\section{Introduction}
For any affine Kac-Moody algebra $\widehat{g}$, let
$\widetilde{U}(\widehat{g})_{k}$ be the completion of its universal enveloping algebra.  It was conjectured that
$\widetilde{U}(\widehat{g})_{k}$ has a center $Z(\widehat{g})$ at
the level $k=-h^\vee$, and furthermore, $Z(\widehat{g})$ possesses a Poisson structure and is isomorphic to the classical W-algebra
${\cal{W}}({g}^{L})$, and ${g}^L$ is Langlands duality of $g$. This is the so called Drinfeld conjecture. Please see\,\cite{FF92}, for more detail statements on the Drinfeld conjecture and its applications. When $k+h^\vee=0$, the level $k$ is nominated as
critical. By applying the Wakimoto realization of
$\widehat{g}$\,\cite{Wak84, FF88}, one can construct a homomorphism from $Z(\widehat{g})$ to a commutative algebra ${\cal H}(g)$ at the critical level. And the associated map ${\cal{W }}(g^L) \rightarrow
{\cal H}(g)$ is nothing but the Miura transformation, which has been defined for any arbitrary $g$\,\cite{DS85}. In particular, for the
center $Z(\widehat{sl_2})$ of $\widetilde{U}(\widehat{sl_2})_{k}$ at the critical level $k=-2$, it is generated by the Sugawara operator
and isomorphic to the algebra ${\cal{W}}(sl_2)$, which is also called the classical Virasoro algebra. Later, E.Frenkel and N.Reshetikhin\,\cite{FR96} generalized the above results to the quantum affine algebra $U_q(\widehat{sl_2})$. Using the Wakimoto
realization of $U_q(\widehat{sl_2})$ given\,\cite{AOS94}, they constructed a homomorphism from the center $Z_q(\widehat{sl_2})$ to
a Poisson algebra  ${\cal H}_q(sl_2)$, which is the q-analogue of the Miura transformation. While, in \,\cite{D98}, the authors constructed the $\hbar$-deformed Miura map and the corresponding deformed Virasoro algebra by considering the Wakimoto realization of
the Sugawara operator for the Yangian double with center
$DY_{\hbar}(sl_2)_k$ at the critical level $k=-2$. These results showed that the Drinfeld conjecture held for the quantum affine algebras and the Yangian double with center. And they have many applications, such as, they gave a new interpretation of the Bethe
ansatz in the Gaudin models of statistical mechanics, which allowed us to relate the Bethe ansatz approach to the geometric Langlands correspondence\,\cite{FFR94, F94}. Recently, in\,\cite{FH10}, the authors applied the above results to study the Langlands duality for
representations of quantum groups.

Elliptic quantum groups were introduced to study the infinite dimensional symmetries of the elliptic face models in statistical mechanics\,\cite{Fe94, Fr97, EF98}. They could be considered as the elliptic deformation of quantum affine algebras. For example, on one
hand, the elliptic quantum algebra $U_{q,p}(\widehat{sl_2})$ degenerates to the quantum affine algebra $U_{q}(\widehat{sl_2})$,
as the parameter $p \rightarrow 0$\,\cite{JKOS991}; on the other hand, $U_{q,p}(\widehat{sl_2})$ can be obtained by twisting the Drinfeld realization of $U_{q}(\widehat{sl_2})$\,\cite{JKOS992}. As a result, it is interesting to ask whether the constructions worked
in the Kac-Moody algebra cases and the $q$-(or $\hbar$)-deformed cases, also work in the elliptic case, and write out explicitly the central elements of the completion of elliptic quantum algebra. It is believed that the statement will be true. For its complexity, the corresponding results are not so straightforward as the previously known cases, so they haven't been given explicitly. This is the goal of this paper.

In\,\cite{JKOS992, K98}, the authors constructed the free field realizations of $U_{q,p}(\widehat{sl_2})$ for any given level $k$. But these results could not be generalized to the higher rank case
$U_{q,p}(\widehat{sl_N})$, whose free field realization with the level $k=1$ was given by\,\cite{KK02}. For arbitrary level, we solved this problem using a different approach\,\cite{CD08, CD09}. Here we will
apply the realization given in \,\cite{CD08} to construct the central elements of the elliptic algebra. We also hope that this construction can be generalized to the higher rank case by applying the results of\,\cite{CD09}.

This paper is organized as follows. In section \ref{Alg}, we will
recall the definition of the elliptic quantum algebra
$E_{q,p}(\widehat{sl_2})$ generated by twisting the generating
currents of $U_{q}(\widehat{sl_2})$. In section \ref{Rep}, for any
given level $k$, we construct the free fields realization of
$E_{q,p}(\widehat{sl_2})$. In section \ref{Cen}, the central
elements of $E_{q,p}(\widehat{sl_2})$ at the critical level $k=-2$
are given. In terms of the free fields, we obtain the same
expression of the central elements as the one of
$U_{q}(\widehat{sl_2})_{k=-2}$. In terms of the free field
realization,there exists a homomorphism from the center
$Z_{q,p}(\widehat{sl_2})$ to a commutative algebra
${\cal{H}}_{q}(sl_2)$. In some sense, it shows that the Drinfeld
conjecture holds in the elliptic case. Throughout this paper, let
$q\in \mathbb{C}$ with $q\neq 0,\ |q |< 1$.

\section{The elliptic algebra $E_{q,p}(\widehat{sl_2})_{k}$}\label{Alg}
In this section, we first review the Drinfeld realization of the quantum affine algebra $U_q(\widehat{sl_2})_{k}$. It is an associative algebra over $\mathbb{C}$ with Drinfeld generators
$E^{\pm}_{n}\,(n\in \mathbb{Z})$, $H_{n}\,(n\in \mathbb{Z}_{\neq
0})$, invertible $q^{h}$ and $c$, which satisfy the following relations \cite{Dri87}:
\beqas && c: central\ \ \ element\\
&&[h, H_n]=0, \ \ \ \
q^{h} E_n^{\pm}q^{-h}=q^{\pm 2}E_n^{\pm}\\
&&[H_n, H_m]=\frac{[2n][cn]}{n}\delta_{n+m, 0},\\
&&[H_n,
E_m^{\pm}]=\pm\frac{[2n]}{n}q^{\mp\frac{c}{2}n}E_{m+n}^{\pm},\\
&&[E_n^{+},
E_m^{-}]=\frac{1}{q-q^{-1}}\big(q^{\frac{c}{2}(n-m)}\psi_{+,
n-m}-q^{-\frac{c}{2}(n-m)}\psi_{-, n-m}\big),\\
&&[E_{n+1}^{\pm}, E_m^{\pm}]_{q^{\pm 2}}+[E_{m+1}^{\pm},
E_n^{\pm}]_{q^{\pm 2}}=0, \eeqas
where $\psi_{\pm, n}$ are defined by
$$\sum_{n\in{\mathbb{Z}}}\psi_{\pm, n}z^{-n}=q^{\pm h}\exp\Big(\pm(q-q^{-1})\sum_{\pm n>0}H_nz^{-n}\Big),$$ and the
symbol $[A,B]_{x}$ for $x\in {\mathbb{C}}$ denotes $AB-xBA$. In terms of the generating functions $K^\pm(z)$, $E^{\pm}(z)$ given by
\beqas &&K^{\pm}(z)=q^{\pm\frac{h}{2}}\exp\Big(\pm(q-q^{-1})\sum_{n>0}\frac{[n]}{[2n]}q^{\mp n}H_{\mp n}z^{\pm n}\Big),\\
&&E^{\pm}(z)=\sum_{n\in{\mathbb{Z}}}E_n^{\pm}z^{-n-1}, \eeqas the
above defining relations can be recast as \beqa
&&K^\pm(z) K^\pm(w)= K^\pm(w) K^\pm(z),\label{kk}\\
&&K^-(z) K^+(w) = \frac{f_{q}(\frac{w}{z}q^{-k})}{f_{q}(\frac{w}{z}q^k)}
K^+(w) K^-(z),\\
&&K^\pm(z) E^{+}(w) = \frac{zq^{\mp\frac{k}{2}-1} - wq}{zq^{\mp\frac{k}{2}} -
w} E^{+}(w) K^\pm(z),\\
&&K^\pm(z) E^{-}(w) = \frac{zq^{\pm\frac{k}{2}} - w}{zq^{\pm\frac{k}{2}-1} -
wq} E^{-}(w) K^\pm(z),\\
&&E^{+}(z) E^{+}(w) = \frac{zq^2-w}{z-wq^2} E^{+}(w) E^{+}(z),\\
&&E^{-}(z) E^{-}(w) = \frac{z-wq^2}{zq^2-w} E^{-}(w) E^{-}(z),\\
&&[ E^{+}(z),E^{-}(w)] =  \frac{1}{(q-q^{-1})zw} \Big( \delta \left(\frac{w}{z}
q^k \right) K^{-}(q^{\frac{k}{2}+2}w)^{-1}K^{-}(q^{\frac{k}{2}}w)^{-1} \nonumber \\
&&\hskip5cm - \delta
\left(\frac{w}{z} q^{-k} \right) K^{+}(q^{-\frac{k}{2}+2}w)^{-1}K^{+}(q^{-\frac{k}{2}}w)^{-1} \Big),\label{ee}
\eeqa
in which the symbol $f_{q}(x)$ is given by
$$f_{q}(x) = \frac{(x;q^4)_{\infty}(xq^4;q^4)_{\infty}}{(xq^2;q^4)_{\infty}^2},$$
with
$$(a;b)_{\infty} = \prod_{n=0}^\infty (1-ab^n);$$
and
$$\delta(x) = \sum_{m\in\mathbb{Z}} x^m.$$

{\noindent}Let the parameters $p$, $p^*$ be
$$p=q^{2r},\ \ p^*=q^{2r^{*}}=pq^{-2c}\ \ \ (r^*=r-c; r,r^{*}\in \mathbb{R}_{> 0})$$
where $c$ is the central element of $U_q(\widehat{sl_2})$. we introduce some dressing currents $V^{\pm}(z; r,r^*)$ and $D^{\pm}(z; r,r^*)$ in $U_{q}(\widehat{sl_{2}})$ depending on the parameters $r$ and $r^*$:
\beqa
&&V^{+}(z; r, r^*)=\exp\Big(-\sum_{n>0}\frac{[n]}{[r^{*}n][2n]}H_{-n}q^{(r^*-1)n}z^n\Big),\label{V}\\
&&V^{-}(z; r, r^*)=\exp\Big(\sum_{n>0}\frac{[n]}{[rn][2n]}H_{n}q^{(r+1)n}z^{-n}\Big),\\
&&D^{+}(z;r, r^*)=\exp\Big(\sum_{n>0}\frac{1}{[r^*n]}H_{-n}q^{(r^{*}+\frac{c}{2})n}z^n\Big),\label{D+}
\\
&&D^{-}(z;r,r^*)=\exp\Big(-\sum_{n>0}\frac{1}{[rn]}H_{n}q^{(r-\frac{c}{2})n}z^{-n}\Big)\label{D}
\eeqa then using them to twist the generating currents $K^{\pm}(z)$
and $E^{\pm}(z)$ of $U_q(\widehat{sl_2})$, we have
{\defn The
twisting currents of the algebra $U_q(\widehat{sl_2})$ are given by
\beqa
&&k^{+}(z;\,r, r^*)=V^{+}(z; r, r^*)K^{+}(z)V^{-}(q^{k}z; r, r^*),\label{k}\\
&&k^{-}(z;\,r, r^*)=V^{+}(q^{k}z; r, r^*)K^{-}(z)V^{-}(z; r, r^*),\\
&&e(z;\, r, r^*)=D^{+}(z; r, r^*)E^{+}(z),\\
&&f(z;\, r, r^*)=E^{-}(z)D^{-}(z; r, r^*)\label{f}. \eeqa} For
brevity, here we just denote them as $k^{\pm}(z;\,p)$, $e(z;\,p)$
and $f(z;\,p)$. Applying the commutation relations
(\ref{kk})-(\ref{ee}), we can prove
{\prop The currents
$k^{\pm}(z;\,p)$, $e(z;\,p)$ and $f(z;\,p)$ satisfy \beqa
&&k^{\pm}(z;\,p)k^{\pm}(w;\,p)=k^{\pm}(w;\,p)k^{\pm}(z;\,p),\label{kkp}\\
&&k^{-}(z;\,p)k^{+}(w;\,p)=\frac{F_{q,\,p}(\frac{w}{z}q^{-k})F_{q,\,p}(\frac{z}{w}q^{k}pp^{*})}
{F_{q,\,p}(\frac{w}{z}q^{k})F_{q,\,p}(\frac{z}{w}q^{-k}pp^{*})}k^{+}(w;\,p)k^{-}(z;\,p),\\
&&k^{\pm}(z;\,p)e(w;\,p)=q\,\cdot\,\frac{\Theta_{p^*}(q^{\mp
\frac{k}{2}-2}\frac{z}{w})} {\Theta_{p^*}(q^{\mp
\frac{k}{2}}\frac{z}{w})}e(w;\,p)k^{\pm}(z;\,p),\eeqa\beqa
&&k^{\pm}(z;\,p)f(w;\,p)=q^{-1}\,\cdot\,\frac{\Theta_{p}(q^{\pm
\frac{k}{2}}\frac{z}{w})}
{\Theta_{p}(q^{\pm \frac{k}{2}-2}\frac{z}{w})}f(w;\,p)k^{\pm}(z;\,p),\\
&&e(z;\,p)e(w;\,p)=q^{-2}\,\cdot\,\frac{\Theta_{p^*}(q^2\frac{z}{w})}
{\Theta_{p^*}(q^{-2}\frac{z}{w})}e(w;\,p)e(z;\,p),\\
&&f(z;\,p)f(w;\,p)=q^2\,\cdot\,\frac{\Theta_p(q^{-2}\frac{z}{w})}
{\Theta_p(q^2\frac{z}{w})}f(w;\,p)f(z;\,p),\\
&&[e(z;\,p), f(w;\,p)]=\frac{1}{(q-q^{-1})zw}\Big(\delta(q^{-c}\frac{z}{w})\Psi
^{+}(q^{\frac{c}{2}}w;\,p)\nonumber\\
&&\hskip7cm -\delta(q^{c}\frac{z}{w})\Psi^{-}(q^{-\frac{c}{2}}w;\,p)\Big)\label{efp}
\eeqa
where the currents $\Psi^{\pm}(z;\,p)$ are given by
$$\Psi^{+}(z;\,p)=:k^{-}(zq^2;\,p)^{-1}k^{-}(z;\,p)^{-1}:$$
$$\Psi^{-}(z;\,p)=:k^{+}(zq^2;\,p)^{-1}k^{+}(z;\,p)^{-1}:$$
and we use the elliptic theta function $\Theta_{t}(z)$
defined for $t\in \mathbb{C}$:
\beqas &&\Theta_{t}(z)=(z;\,t)_{\infty}(tz^{-1};\,t)_{\infty}(t;\,t)_{\infty},
\\
&&(z; t_1,\cdots, t_k)_{\infty}=\prod_{n_1, \cdots,
n_k\geq0}(1-zt_1^{n_1}\cdots t_k^{n_k}) \eeqas
\no and
$$F_{q,\,p}(x)=\frac{(x;\,q^{4},p,p^*)_{\infty}(xq^{4};\,q^{4},p,p^*)_{\infty}}{(x q^{2};\,q^{4},p,p^*)_{\infty}^{2}}
.$$}

It should be remarked that the factor $F_{q,\,p}(x)$ depends on the two parameters $q$ and $p$. It is two-parameter generalization of the scalar factor $f_{q}(x)$ appeared in\,\cite{FR96}. It has close relationship with the algebraic structure defined by (\ref{kkp})-(\ref{efp}), just like the roles played by the function $f_{q}(x)$ given above. And it is also related to the crossing-symmetry property of the elliptic type R-matrix. Furthermore, as $p\rightarrow 0$,
$$F_{q,\,p}(x)\rightarrow f_{q}(x),$$
then the relations (\ref{kkp})-(\ref{efp}) degenerate to
(\ref{kk})-(\ref{ee}). Here we just denote the algebra generated by
the twisting currents $k^{\pm}(z;\,p)$, $e(z;\,p)$,$f(z;\,p)$ and
the central element $c$ with the defining relations
(\ref{kkp})-(\ref{efp}) as $E_{q,p}(\widehat{sl_{2}})$.

\section{The Wakimoto realization of $E_{q,p}(\widehat{sl_{2}})$}\label{Rep}
In this section, we will first recall the Wakimoto realization of
$U_q(\widehat{sl_2})$ for any given level \cite{AOS94}. Then by
introducing some twisting currents, we construct the free fields
realization of the algebra $E_{q,p}(\widehat{sl_{2}})$. There are
three free bosons \beqas
&&[a_n,a_m]=\frac{[(k+2)n][2n]}{n}\delta_{n+m, 0},\ \ \ [p_a,
q_a]=2(k+2),\\
&&[b_n, b_m]=-\frac{[n]^2}{n}\delta_{n+m, 0},\ \ \ [p_b, q_b]=-1,\\
&&[c_n, c_m]=\frac{[n]^2}{n}\delta_{n+m, 0},\ \ \ [p_c, q_c]=1 \eeqas
which generate a quantum Heisenberg algebra ${\cal H}_{q, k}$. Using them, one can introduce some
free boson fields
\beqas
&&A_{\pm}(z)=\pm\big((q-q^{-1})\sum_{n>0}\frac{[n]}{[2n]}a_{\pm n}z^{\mp n}+\frac{p_{a}}{2}\ln q\big),\\
&&a(z;\alpha)=-\sum_{n\neq 0}\frac{a_n}{[n]}q^{-\alpha n}z^{-n}+q_{a}+p_{a}\ln z,\\
&&a_{\pm}(z)=\pm\big((q-q^{-1})\sum_{n> 0}a_{\pm n}z^{\mp n}+p_{a}\ln q\big)
\eeqas
where $\alpha\in {\mathbb{C}}$. For brevity, let $a(z; 0)=a(z)$. Similarly, we can define other free boson fields $b(z; \alpha)$, $b_{\pm}(z)$ and $c(z; \alpha)$, $c_{\pm}(z)$. And the normal ordering $:\ :$ is defined by moving $a_{n}\,(n>0)$, $p_{a}$ to right and moving $a_{n}\,(n<0)$, $q_{a}$ to left. For example,
$$:\,\exp(a(z;\,\alpha))\,:=\exp\big(-\sum_{n<0}\frac{a_{n}}{[n]}(q^{-\alpha}z)^{-n}\big)
e^{q_{a}}z^{p_{a}}\exp\big(-\sum_{n>0}\frac{a_{n}}{[n]}(q^{\alpha}z)^{-n}\big).$$

{\noindent}By\,\cite{AOS94}, we know that there exists a
homomorphism $h_{q, k}$ from the algebra
$U_{q}(\widehat{sl_{2}})_{k}$ to the algebra ${\cal H}_{q, k}$ given
by \beqas
&&h_{q,k}(K^{+}(z))={A_{-}(zq^{-2})}^{-1}\exp\big(-b_{-}(zq^{-\frac{k}{2}-2})\big),\\
&&h_{q,k}(K^{-}(z))={A_{+}(z)}^{-1}\exp \big(-b_{+}(zq^{\frac{k}{2}})\big),\\
&&h_{q,k}(E^{+}(z))=:\exp\big(b_{-}(z)-(b+c)(zq^{-1})\big):\\
&&\hskip2.7cm -:\exp\big(b_{+}(z)-(b+c)(z q)\big): \\
&&h_{q,k}(E^{-}(z))=A_{+}(zq^{\frac{k}{2}})A_{+}(zq^{\frac{k}{2}+2}):\exp\big (b_{+}(zq^{k+2})+(b+c)(zq^{k+1})\big):\\
&&\hskip2cm
-\,A_{-}(zq^{-\frac{k}{2}})A_{+}(zq^{-\frac{k}{2}-2}):\exp\big
(b_{-}(zq^{-k-2})+(b+c)(zq^{-k-1})\big): \eeqas which is called the
Wakimoto realization of $U_q(\widehat{sl_2})$ with level $k$. In
particular, in terms of the free bosons, the generators
$H_{n}\,(n\in \mathbb{Z}-\{0\})$ are represented by
$$h_{q,k}(H_{n})=a_{n}q^{-\mid n \mid}+b_{n}(q^{-\frac{k}{2}\mid n \mid}+q^{-(\frac{k}{2}+2)\mid n \mid}),$$
then by (\ref{V})-(\ref{D}), we obtain the free boson realization of
the twisting currents $V^{\pm}(z; r,r^*)$ and $D^{\pm}(z; r, r^*)$.
Furthermore, applying them to the relations (\ref{k})-(\ref{f}), we
obtain the free bosons realization of the algebra
$E_{q,p}(\widehat{sl_{2}})$ with level $c=k$. More precise, we have
another homomorphism $h_{q, p; k}$ from the algebra
$E_{q,p}(\widehat{sl_{2}})$ to the Heisenberg algebra ${\cal
H}_{q,k}$. For simplicity, we will just use the same notations for
the generating currents and the corresponding images of them under
the homomorphism $h_{q,p; k}$.

It should be remarked that when $k\neq -2$, the homomorphism $h_{q,p; k}$ provides representations of $E_{q,p}(\widehat{sl_{2}})$ in the Fock representation of the Heisenberg algebra ${\cal H}_{q,k}$ similarly with the discussion in \cite{AOS94}. These representations have one parameter the action of $p_{a}$ on the
highest weight vector. When $k=-2$, the generators $a_n$ commute among themselves and generate a commutative algebra ${\cal H}_q(sl_2)$. Therefore representations of
$E_{q,p}(\widehat{sl_{2}})$ at the critical level $k=-2$ can be realized via $h_{q,p; k}$ in a smaller space: the tensor product of the Fock representation of the subalgebra of ${\cal H}_{q,-2}$ generated by $b_n, c_n, n\in\mathbb{Z}$, and a one-dimensional
representation of ${\cal H}_q(sl_2)$. In next section, we will apply the representation at the critical level to construct the central element for the elliptic algebra $E_{q,p}(\widehat{sl_{2}})$.

\section{Central elements of $E_{q,p}(\widehat{sl_{2}})$}\label{Cen}
In this section we will apply the Wakimoto realization of
$E_{q,p}(\widehat{sl_{2}})$ at the critical level to construct the
center of $E_{q,p}(\widehat{sl_{2}})$. We first recall the
Ding-Frenkel correspondence of the algebra
$U_{q}(\widehat{sl_{2}})_{k}$\,\cite{DF93}, which was used to
construct the isomorphism between the Drinfeld realization and the
RS realization of $U_{q}(\widehat{sl_{2}})$. More precise, the
generating functions $L^{\pm}(z)$ of the RS realization can be
decomposed as:
$$L^\pm(z) =
\begin{pmatrix}
1 & 0 \\
e^\pm(z) & 1
\end{pmatrix}
\begin{pmatrix}
K^\pm(z) & 0 \\
0 & {K^\pm(zq^{2})}^{-1}
\end{pmatrix}
\begin{pmatrix}
1 & f^\pm(z) \\
0 & 1
\end{pmatrix}$$
in which the notations $e^{\pm}(z)$ and $f^{\pm}(z)$ are called the half currents of $E^{+}(z)$ and $E^{-}(z)$, since they satisfy
\beqas
&&E^{+}(z)\,=\,e^{+}(zq^{\frac{k}{2}})-e^{-}(zq^{-\frac{k}{2}}),\\
&&E^{-}(z)\,=\,f^{+}(zq^{-\frac{k}{2}})-f^{-}(zq^{\frac{k}{2}}).
\eeqas

{\noindent} Using $D^{\pm}(z;r, r^*)$ defined by
(\ref{D+})-(\ref{D}) to dress the half currents $e^{\pm}(z)$ and
$f^{\pm}(z)$, we define the twisting of above half currents as \beqas
&&e^{+}(z; r, r^*)\,=\,D^{+}(zq^{-\frac{k}{2}}; r, r^*)e^{+}(z),\\
&&e^{-}(z; r, r^*)\,=\,D^{+}(zq^{\frac{k}{2}}; r, r^*)e^{+}(z),\\
&&f^{+}(z; r, r^*)\,=\,f^{+}(z)D^{-}(zq^{\frac{k}{2}}; r, r^*),\\
&&f^{-}(z; r, r^*)\,=\,f^{+}(z)D^{-}(zq^{-\frac{k}{2}}; r, r^*);
\eeqas respectively, and we will simply denote them as $e^{\pm}(z;\,p)$ and
$f^{\pm}(z;\,p)$; then by direct computation, we obtain \beqas
&&e(z;\,p)=e^{+}(zq^{\frac{k}{2}};\,p)-e^{-}(zq^{-\frac{k}{2}};\,p),\\
&&f(z;\,p)=f^{+}(zq^{-\frac{k}{2}};\,p)-f^{-}(zq^{\frac{k}{2}};\,p).
\eeqas
For any $k$, if we further introduce the generating function
\beqas
&&l(z;\,p)\,=\,q^{-1}:k^{+}(zq^{-\frac{k}{2}};\,p){k^{-}(zq^{\frac{k}{2}};\,p)}^{-1}:
+q:k^{-}(zq^{\frac{k}{2}+2};\,p){k^{+}(zq^{-\frac{k}{2}+2};\,p)}^{-1}:\\
&&\hskip3cm +k^{+}(zq^{-\frac{k}{2}};\,p):e(z;\,p)f(z;\,p):k^{-}(zq^{\frac{k}{2}+2};\,p)
\eeqas
where
$$:e(z;\,p) f(z;\,p):\,=\,e^{+}(zq^{\frac{k}{2}};\,p)f(z;\,p)-f(z;\,p)e^{-}(zq^{-\frac{k}{2}};\,p),$$
then we can prove the following theorem {\thm When $k=-2$, the
coefficients of the generating function \beqas
&&l(z;\,p)\,=\,q^{-1}:k^{+}(zq;\,p){k^{-}(zq^{-1};\,p)}^{-1}:+\,q:k^{-}(zq;\,p){k^{+}(zq^{3};\,p)}^{-1}:\\
&&\hskip3cm +k^{+}(zq;\,p):e(z;\,p)f(z;\,p):k^{-}(zq;\,p) \eeqas are
the central elements of $E_{q,p}(\widehat{sl_{2}})_{k}$. }

The above identity could be considered as a transfermation from transfer matrix to quantum trace
(T-Q transfermation, for short) depending on two parameters. Here we will briefly present how to prove Thm.1 by applying the Wakimoto
realization of the algebra $E_{q,p}(\widehat{sl_{2}})$. In terms of the free bosons fields, we first consider the normally ordered product $:e(z;\,p) f(z;\,p):$. On one hand, using the Wakimoto realization of $D^{\pm}(z;\,r,r^*)$ and $E^{\pm}(z)$, we have
$$e(w;\,p)f(z;\,p)=D^{+}(w;\,r,r^*)E^{+}(w)E^{-}(z)D^{-}(z;\,r,r^*),$$
in which
\beqas
&E^{+}(w)E^{-}(z)=q A_+(zq^{-1}) A_+(zq)e^{b_-(w)}
e^{(b+c)(zq^{-1}) - (b+c)(wq^{-1})} e^{b_+(z)}\\
&+ q^{-1}A_-(zq^{-1})A_-(zq)e^{(b_-(z))}e^{(b+c)(zq) -
(b+c)(wq)} e^{b_+(w)} \\
&- \frac{w-z}{wq-zq^{-1}}A_+(zq^{-1})
A_+(zq)e^{(b+c)(zq^{-1}) - (b+c)(wq)} e^{b_+(z) + b_+(w)}\\
&- \frac{w-z}{wq^{-1}-zq}A_-(zq^{-1}) A_-(zq)e^{b_-(z) +
b_-(w)} e^{(b+c)(zq)-(b+c)(wq^{-1})}.
\eeqas
Similarly with $E^{+}(w)E^{-}(z)$ discussed in\,\cite{FR96}, $e(w;\,p)f(z;\,p)$ makes sense in the region $|w|>|z|$, $|w|>q^{2}|z|$ and $|w|>q^{-2}|z|$. On the other hand, we have
\beqas
f(z;\,p)e(w;\,p)&=&E^{-}(z)D^{-}(z;\,r,r^*)D^{+}(w;\,r,r^*)E^{+}(w)\\
&=&c_{1}E^{-}(z)D^{+}(w;\,r,r^*)D^{-}(z;\,r,r^*)E^{+}(w)\\
&=&c_{1}c_{2}D^{+}(w;\,r,r^*)E^{-}(z)D^{-}(z;\,r,r^*)E^{+}(w)\\
&=&c_{1}c_{2}c_{3}D^{+}(w;\,r,r^*)E^{-}(z)E^{+}(w)D^{-}(z;\,r,r^*)\\
&=&D^{+}(w;\,r,r^*)E^{-}(z)E^{+}(w)D^{-}(z;\,r,r^*),
\eeqas
in which the coefficients are given by
\beqas
&&c_{1}=\frac{\big(pq^{4}\frac{w}{z};\,p\big)_{\infty}\big(p^{*}q^{-4}\frac{w}{z};\,p^*\big)_{\infty}}
{\big(pq^{-4}\frac{w}{z};\,p\big)_{\infty}\big(p^{*}\frac{w}{z};\,p^*\big)_{\infty}},\\
&&c_{2}=\frac{\big(p^{*}\frac{w}{z};\,p^*\big)_{\infty}}{\big(p^{*}q^{-4}\frac{w}{z};\,p^*\big)_{\infty}},\\
&&c_{3}=\frac{\big(pq^{-4}\frac{w}{z};\,p\big)_{\infty}}{\big(pq^{4}\frac{w}{z};\,p\big)_{\infty}}
\eeqas
and $c_{1}c_{2}c_{3}=1$, then $f(z;\,p)e(w;\,p)$ has the same formula as $e(w;\,p)f(z;\,p)$, but makes sense in the region $|w|<|z|$, $|w|<q^{2}|z|$ and $|w|<q^{-2}|z|$. Therefore, we can write $:e(z;\,p) f(z;\,p):$ as
$$:e(z;\,p) f(z;\,p):=\int_{C_{R}}\frac{e(w;\,p)f(z;\,p)}{w-z}dw-\int_{C_{r}}\frac{f(z;\,p)e(w;\,p)}{w-z}dw$$
where the notations $C_{R}$ and $C_{r}$ are circles around the origin of radii $R>|w|$ and $r<|w|$ respectively, and here integrals are the contours on the $w$ plane surrounding the points $z,\, zq^2, \,zq^{-2}$. As a result, we obtain the expression
\beqas
&:e(z;\,p) f(z;\,p): =q D^{+}(z;\,r,r^*)A_+(zq^{-1}) A_+(zq) e^{b_-(z)} e^{b_+(z)}D^{-}(z;\,r,r^*)\\
&+ q^{-1} D^{+}(z;\,r,r^*)A_-(zq^{-1}) A_-(zq)e^{b_-(z)} e^{b_+(z)}D^{-}(z;\,r,r^*) \\
&- q^{-1} D^{+}(z;\,r,r^*)A_+(zq^{-1}) A_+(zq) e^{b_+(zq^{-2}) + b_+(z)}D^{-}(z;\,r,r^*)\\
&-q D^{+}(z;\,r,r^*)A_-(zq^{-1}) A_-(zq) e^{b_-(zq^2)+b_-(z)}D^{-}(z;\,r,r^*).
\eeqas

{\noindent}Applying the free fields realization of $k^{\pm}(z;\,p)$ and $D^{\pm}(z;\,r,r^*)$ given in above section, we have
\beqas
&&k^+(zq;\,p) :e(z;\,p) f(z;\,p): k^-(zq;\,p)\\
&&=q^{-1} A_-(zq)A_+(zq)^{-1}+ q A_-(zq^{-1})^{-1} A_+(zq^{-1})\\
&&- q^{-1}W_{1}(q^{-1}z;\,r,r^*)A_-(zq^{-1})^{-1}
A_+(zq^{-1})W_{2}(q^{-1}z;\,r,r^*)^{-1}e^{-b_-(z)} e^{b_+(zq^{-2})} \\
&&- q W_{1}(qz;\,r,r^*)^{-1}A_-(zq)A_+(zq)^{-1}W_{2}(qz;\,r,r^*) e^{b_-(zq^2)} e^{-b_+(z)}.
\eeqas
where
\beqas
&&W_{1}(z;\,r,r^*)=\exp\Big(-(q-q^{-1})\sum_{n>0}\frac{[n]}{[r^*n]}a_{-n}q^{rn}z^n\Big),\\
&&W_{2}(z;\,r,r^*)=\exp\Big((q-q^{-1})\sum_{n>0}\frac{[2n]}{[r^*n]}b_{-n}q^{(r+1)n}z^n\Big).
\eeqas
Moreover,
\begin{align*}
q^{-1}k^{+}(zq;\,p){k^{-}(zq^{-1};\,p)}^{-1}=&q^{-1}W_{1}(q^{-1}z;\,r,r^*)A_-(zq^{-1})^{-1}
A_+(zq^{-1})\\
& \times W_{2}(q^{-1}z;\,r,r^*)^{-1} e^{-b_-(z)}e^{b_+(zq^{-2})},
\end{align*}
and
\begin{align*}
q{k^{+}(zq^{3};\,p)}^{-1}{k^{-}(zq^{-1};\,p)}=&qW_{1}(qz;\,r,r^*)^{-1}A_-(zq)A_+(zq)^{-1}\\
& \times W_{2}(qz;\,r,r^*) e^{b_-(zq^2)} e^{-b_+(z)};
\end{align*}
then under the homomorphism $h_{q,p; k}$ with $k=-2$, the current
$l(z;\,p)$ can be represented as
$$l(z;\,p)=q^{-1}A_{-}(zq)A{+}(zq)^{-1}+qA_{-}(zq^{-1})^{-1}A_{+}(zq^{-1}).$$
It says that in terms of the free fields, the Cartan parts are decoupled with the off-diagonal ones. Then we can further show that under the Wakimoto representation $h_{q,p; k}$ with the critical
level $k=-2$, $l(z;\,p)$ commutes with the twisting currents
$k^{\pm}(z;\,p)$, $e(z;\,p)$ and $f(z;\,p)$, i.e., the Fourier
coefficients of $l(z;\,p)$ are the central elements of the elliptic
algebra $E_{q,p}(\widehat{sl_2})$.

It should also be noted that although the expression for $l(z;\,p)$
in terms of free fields is the same as the one for the trigonometric
case $U_{q}(\widehat{sl_2})_{k=-2}$\,\cite{FR96}, their algebraic
expressions are different from each other. In fact, we obtain the
elliptic algebra $E_{q,p}(\widehat{sl_2})$ by twisting the quantum
affine algebra $U_{q}(\widehat{sl_2})$. The twisting currents map a
Hopf algebra structure of $U_{q}(\widehat{sl_2})$ to a quasi-Hopf
algebra structure of $E_{q,p}(\widehat{sl_2})$\,\cite{JKOS991}. As
what we have clarified in this paper, it doesn't change the center
of them in terms of the free fields.

\section{Discussion}
In this paper, applying the free field realization of
$E_{q,p}(\widehat{sl_2})$, we construct the center of it at the
critical level. We hope that the construction can be generalized to
the higher rank case. It is also interesting to prove the results
without using the free field representation expressions. We hope to
discuss these problems in future.

\vskip 1cm
\section{Acknowledgments}
\vskip 0.5cm

One of the authors (Ding) is financially supported partly by the
Natural Science Foundations of China through the grands No.10931006 and
No.10975180. He would like to
thank the Institute of Mathematical Sciences, the Chinese University of Hong Kong for hospitality, where
part of this work was done.

\end{document}